\documentclass[10pt,superscriptaddress,aps,pra,twocolumn,showpacs,floatfix]{revtex4-2}
\usepackage[utf8]{inputenc}
\usepackage[percent]{overpic}
\usepackage{lipsum}
\usepackage{bm}
\usepackage[bbgreekl]{mathbbol}
\usepackage{color} 
\usepackage{braket}
\usepackage{gensymb}
\usepackage[dvipsnames]{xcolor}
\usepackage{lpic}
\usepackage{physics}
\usepackage{hyperref}
\usepackage{graphicx}
 \hypersetup{
    colorlinks=true,
    linkcolor=blue,
    filecolor=magenta,
    citecolor=blue,
    urlcolor=cyan,
}

\newcommand{\orcid}[1]{\href{https://orcid.org/#1}{\includegraphics[width=7pt]{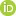}}}

\begin{document}
\preprint{APS/123-QED}

\title{Thermodynamical aspects of optically pumped dense atomic medium}%

\author{A. F. Sousa\orcid{0009-0004-8844-0218}}
\email{aryadine.sousa@ufabc.edu.br}
\affiliation{Centro de Ci\^{e}ncias Naturais e Humanas, Universidade Federal do ABC, Avenida dos Estados 5001, 09210-580 Santo Andr\'e, S\~{a}o Paulo, Brazil}

\author{C. H. S. Vieira\orcid{0000-0001-7809-6215}}
\email{vieira.carlos@ufabc.edu.br}
\affiliation{Centro de Ci\^{e}ncias Naturais e Humanas, Universidade Federal do ABC, Avenida dos Estados 5001, 09210-580 Santo Andr\'e, S\~{a}o Paulo, Brazil}
\affiliation{Department of Physics, State Key Laboratory of Quantum Functional Materials,
and Guangdong Basic Research Center of Excellence for Quantum Science,
Southern University of Science and Technology, Shenzhen 518055, China}

\author{H. M. Florez\orcid{0000-0001-5847-1890}}
\email{hans.m@ufabc.edu.br}
\affiliation{Centro de Ci\^{e}ncias Naturais e Humanas, Universidade Federal do ABC, Avenida dos Estados 5001, 09210-580 Santo Andr\'e, S\~{a}o Paulo, Brazil}


\begin{abstract}
Optically Pumped Magnetometers use light to drive an atomic vapor into a Non-Equilibrium Steady State for sensing. This kind of state is achieved when spin-exchange collisions, together with optical pumping, dominate the relaxation dynamics, redistributing the atomic populations and thereby shaping the steady-state configuration. Despite the rapid advancement of atomic magnetometer technology, a comprehensive thermodynamic analysis of the state preparation is largely unexplored. We apply a thermodynamic framework to alkali atoms in a vapor cell, modeling their interactions with the pump laser and their relaxation via spin-exchange and spin-destruction collisions. We analyze how the pump rate and light polarization determine the non-equilibrium steady state, quantifying irreversibility via entropy production, assessing useful energy via ergotropy, and defining the spin-polarization efficiency. Finally, we establish a connection between metrological performance and the Quantum Fisher Information (QFI), demonstrating that a higher thermodynamic efficiency directly translates into an improved fundamental bound on magnetometer sensitivity. These results provide insights for optimizing state preparation in quantum sensors.
\end{abstract}


\maketitle
\section{Introduction}
High-sensitivity magnetometry has a wide range of applications, from medical imaging \cite{Murzin2020}, such as magnetoencephalography (MEG) \cite{ReviewMEG2022} and magnetocardiography (MCG) \cite{Lin2024MCG}, to geophysical applications, such as geonavigation and mining \cite{LuGEOMAG2023}. Among the available technologies, Optically Pumped Magnetometers (OPMs) have emerged as a promising platform \cite{Kominis2003Nat}, combining high sensitivity, high sample rate, and low operational cost.

An OPM operates through a three-stage process. First, a pump laser is used to polarize an atomic vapor, aligning the atomic spins and driving the system into a Non-Equilibrium Steady State (NESS). The interplay between optical pumping, spin-exchange collisions, and spin destruction relaxation processes governs the state preparation. Second, these polarized spins precess in an external magnetic field at the Larmor frequency, which is directly proportional to the strength of the external field. Finally, a probe laser measures this precession, typically by detecting the rotation of its own polarization, allowing for a precise determination of the magnetic field \cite{Fabricant_2023}. All these stages can affect the device's sensitivity, but the spin-exchange collision plays a particular role.

Seminal works by Happer have described the dynamics of polarized spins in dense media and how the unique relaxation effect due to spin-exchange collisions is inhibited in the Spin Exchange Relaxation Free (SERF) regime \cite{Happer73, Happer77}. This means that light can effectively transfer energy to the atoms despite spin-exchange collisions, making them a resource for sensing.  The approach to this kind of system has focused on spin polarization. However, there is no qualitative description of how light transfers energy to the spins, leading to spin organization, nor is there a figure of merit for how the atoms store that energy.  Pioneering work has been done~\cite{Kominis2024}, connecting the sensitivity limits of the readout stage to concepts such as Landauer's erasure principle. However, this thermodynamic analysis is primarily concerned with the final measurement step. In this context, thermodynamics can provide a more complete description of the optical pumping process by accounting for energy transfer from light to the spin state, spin organization, and the process's irreversibility.

In this work, we extend the thermodynamic analysis of OPMs to the dynamics of the spins for high-density media. We apply a quantum thermodynamic framework~\cite{campbell2025roadma,binder_book,Vinjanampathy,Goold_2016,Deffner_AVS,Vieira2023} to characterize optical pumping as a driven-dissipative process, investigating the trade-off between the thermodynamic cost (entropy production) and the generation of useful energy (ergotropy). Based on this analysis, we quantify the preparation efficiency. Furthermore, we establish a connection to metrological performance via Quantum Fisher Information (QFI)~\cite{Seth_PRL06,Giovannetti2011}. We demonstrate that a more efficiently prepared state yields a higher QFI, thereby improving the fundamental bound on the magnetometer's sensitivity.
The dynamical results presented here are derived from the formalism established in Refs.~\cite{Romalis2002PRL, Kominis2003Nat}. While those studies characterized the spin evolution and relaxation mechanisms, the thermodynamic implications of these dynamics have not been previously explored. 

The paper is organized as follows. In Section~\ref{secII}, we detail our system model, centered on the master equation governing the dynamics of the atomic ensemble. This equation incorporates the key physical interactions: the driving by the pumping laser, spin-exchange collisions, and spin-destruction collisions, along with their numerical solution. In Section~\ref{secIII}, we analyze the ordering of the system induced by the pumping process and investigate the behavior of entropy production. Section~\ref{secIV} introduces the concepts of ergotropy and polarization efficiency, proposing them as a method to analyze the effectiveness of the state preparation stage. In addition, in Section \ref{secV} we discuss the QFI to establish the connection between these thermodynamic figures of merit and the bound sensitivity of the magnetometer. Finally, Section~\ref{secVI} presents our conclusions.
\section{Theoretical Model of the OPM}~\label{secII}
\begin{figure}[b]
\centering
\includegraphics[width=0.7\linewidth, angle=0]{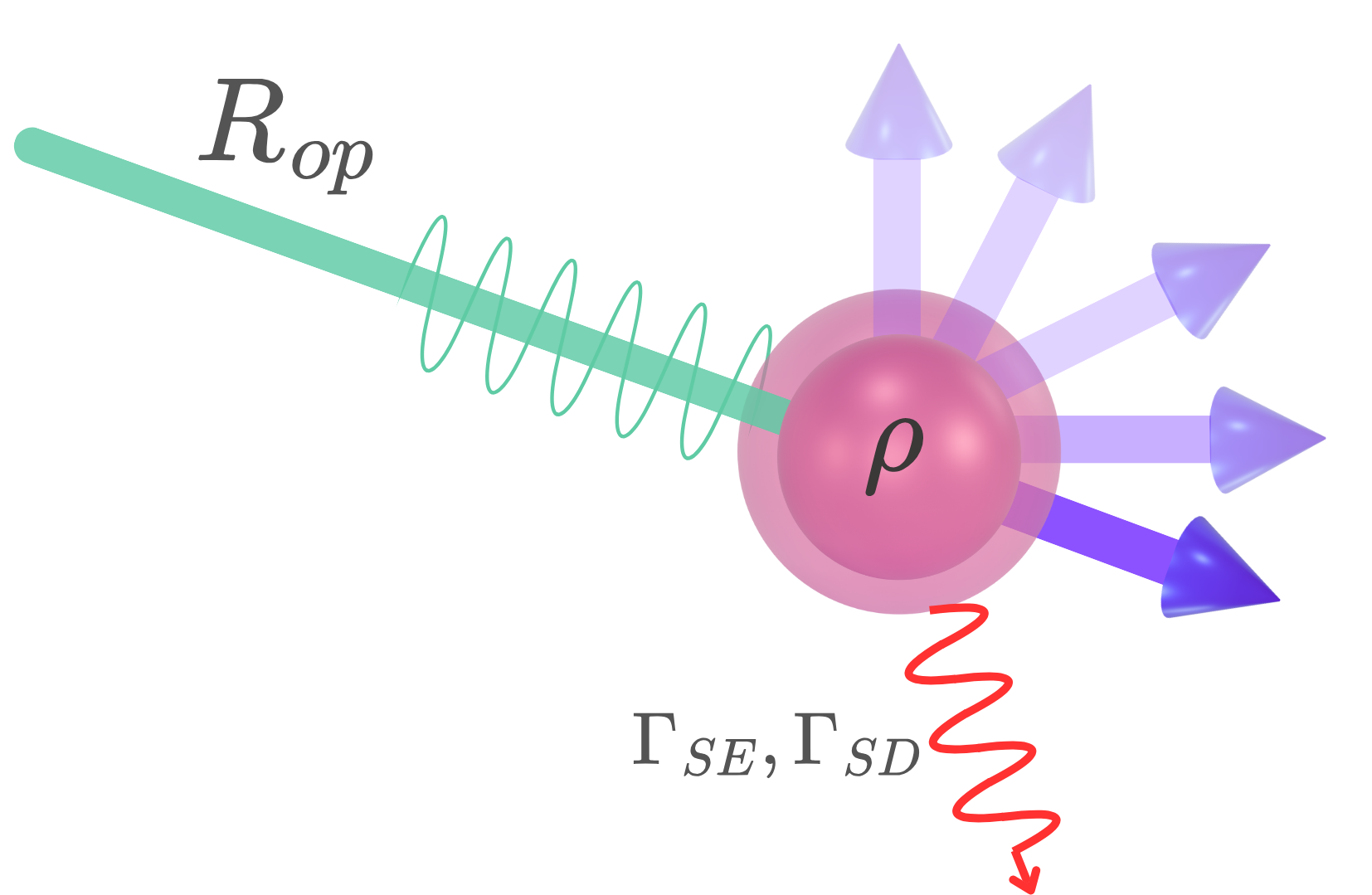}
\caption{An illustration of the optical pumping process. The atomic spin polarization, $\rho$, is driven by the pump laser, $R_{\text{op}}$, and decays due to spin-exchange, $\Gamma_{\text{SE}}$, and spin-destruction, $\Gamma_{\text{SD}}$, processes.}    \label{fig:diagram}
\end{figure}

We analyze the dynamics of an optically pumped ensemble of alkali-metal atoms. The density matrix, $\rho$, describes a single representative atom in the ensemble, yet its dynamics includes interactions with other atoms and with the vapor cell walls. Considering the internal structure, an alkali atom in its electronic ground state ($^2S_{1/2}$) possesses an electronic spin $\mathbf{S}$ (with $S=1/2$) and a nuclear spin $\mathbf{I}$. The coupling between the electronic and nuclear spins gives rise to the total atomic spin operator, $\mathbf{F} = \mathbf{S} + \mathbf{I}$, defining the hyperfine structure of the atom. The internal coherent dynamics are governed by the Hamiltonian $H_0 = A_{\text{hfs}} \mathbf{I} \cdot \mathbf{S}$, where $A_{\text{hfs}}$ is the hyperfine coupling constant\cite{Happer1972}. To operate as an OPM, the atomic ensemble is driven out of thermal equilibrium by a pumping laser at a rate $R_{\text{op}}$. The interaction with the polarization of the light produces orientation of the atomic spin, which subsequently evolves under spin-exchange ($\Gamma_{\text{SE}}$) and spin-destruction ($\Gamma_{\text{SD}}$) processes, as schematically illustrated in Fig.~\ref{fig:diagram}.

The polarization of the laser light is characterized by the photon spin vector $\mathbf{s}$. The complete evolution of the density matrix $\rho$, incorporating optical pumping and collisional relaxation, is described by the master equation~\cite{Happer1972,Happer73,Romalis2002PRL}:
\begin{align}
 \frac{d\rho}{dt} &= \frac{1}{i\hbar}[H_0, \rho]+ R_{\text{op}}\left[\phi(1+2\mathbf{s}\cdot\mathbf{S}) - \rho\right] \nonumber\\
 &+ \Gamma_{\text{SE}}\left[\phi(1+4\langle\mathbf{S}\rangle\cdot\mathbf{S}) - \rho\right] + \Gamma_{\text{SD}}[\phi - \rho]. 
\label{eq:full_master_eq}   
\end{align}
The term proportional to $R_{\text{op}}$ describes how the laser polarizes the electronic spins $\mathbf{S}$, and $\phi =  \rho/4  + \mathbf{S}\rho\mathbf{S}$
represents the nuclear part of the density matrix, with electronic correlations removed. The term $\Gamma_{\text{SE}}$ models the spin-exchange (SE) collisions with other atoms in the vapor cell, while $\Gamma_{\text{SD}}$ represents the spin-destruction (SD) collisions, such as those with the cell walls.

In this work, we focus on the dynamics of Rubidium-87 ($^{87}$Rb), although the analysis presented in the following sections works for any alkali atom. For this isotope, the nuclear spin is $I=3/2$. Consequently, the coupling $\mathbf{F} = \mathbf{S} + \mathbf{I}$ results in two hyperfine ground state manifolds, $F=1$ and $F=2$. The density matrix is then represented in the basis of Zeeman sublevels $|F, m_F\rangle$:
\begin{equation}
\rho=\sum_{F,m_{F},F',m_{F'}}p_{F,m_{F},F',m_{F'}}|F,m_{F}\rangle\langle F',m_{F}'|.
\end{equation}  
Appendix~\ref{apxA} contains additional information on the numerical implementation and the specific parameter values selected for Eq.~(\ref{eq:full_master_eq}). 

\begin{figure}[ht!]
\centering
 \begin{overpic}[width=0.48\textwidth]{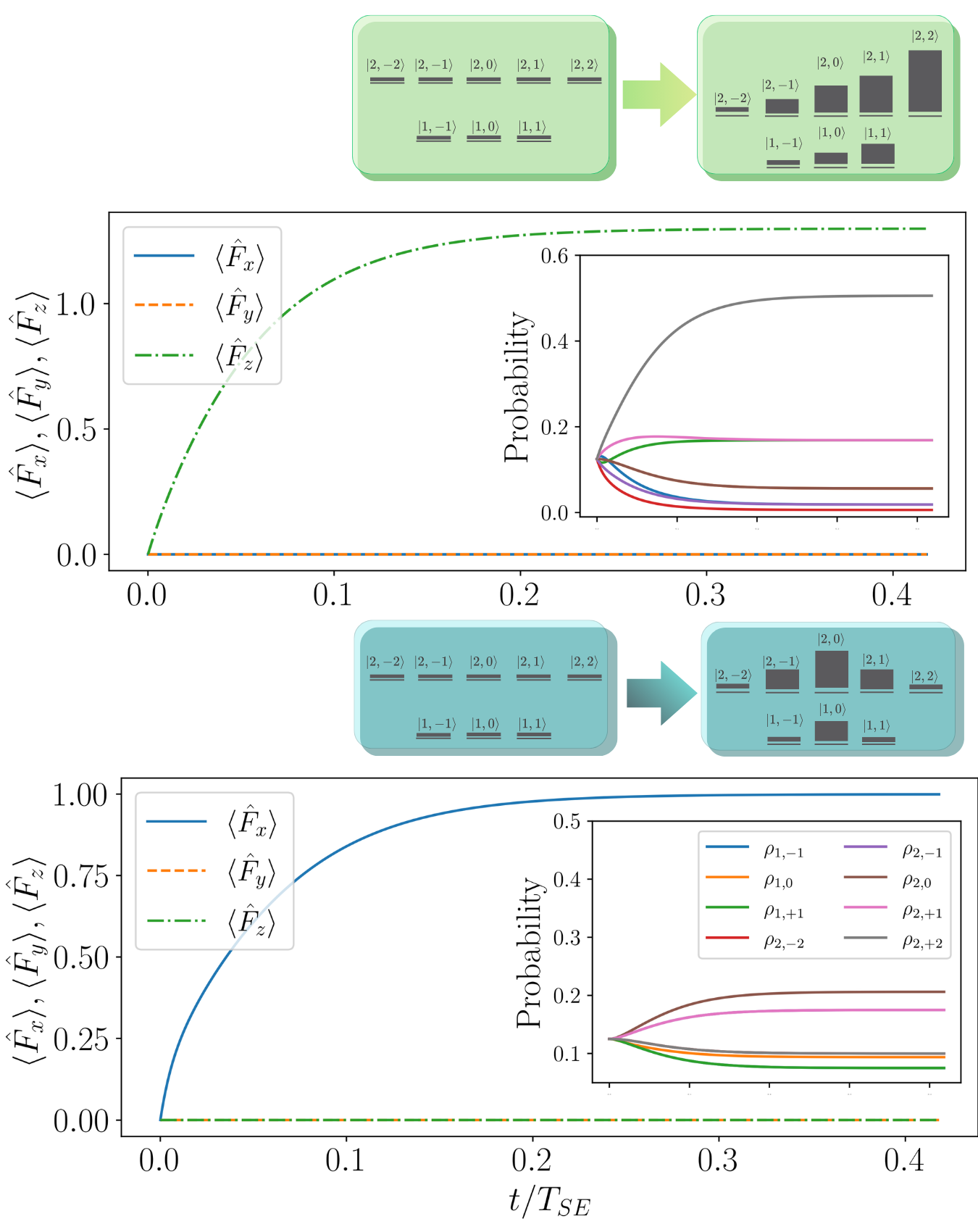}
  \put(0.01,85){(a)}
\put(0.01,45){(b)}
 \end{overpic}
  \caption{Polarization dynamics of an optically pumped $^{87}$Rb ensemble driven by (a) $\hat{z}$-polarized and (b) $\hat{x}$-polarized  light. The schematic diagrams above the plots illustrate the population distributions of the hyperfine Zeeman sublevels $\ket{F,m_F}$ for the initial unpolarized state (left) and the NESS (right). The main plots show the total spin expectation values $\langle\hat{F}_i\rangle$, while the insets show the Zeeman state populations as a function of normalized time $t/T_{\text{SE}}$. The dynamics are solved for a vapor cell (radius $1.5\ \text{cm}$) at $120\ ^{\circ}\text{C}$ with buffer gas pressures of $200\ \text{Torr}$ He and $75\ \text{Torr}$ $\text{N}_2$. The relaxation rates used are $\Gamma_{\text{SD}} \approx 30\ \text{Hz}$ and $\Gamma_{\text{SE}} = 14\ \text{kHz}$. The optical pumping rate is set to $R_{\text{op}} = \Gamma_{\text{SE}}$ with $s=0.5$.}
\label{fig:polarization}
\end{figure}

Figure~\ref{fig:polarization} shows the time evolution of the resulting density matrix $\rho_t$ from Eq.(\ref{eq:full_master_eq}) and the expected value of the total atomic spin, $\langle\mathbf{F}\rangle$. The light polarization dictates the orientation of the atomic polarization. Figure \ref{fig:polarization}(a) shows that pumping along the $\hat{z}$-axis results in a non-zero $\langle\hat{F}_z\rangle$, while Fig. \ref{fig:polarization}(b) shows that pumping along the $\hat{x}$-axis produces a non-zero $\langle\hat{F}_x\rangle$. The insets of Fig.~\ref{fig:polarization} illustrate the dynamics of the state populations. The system begins in a maximally mixed state, evolves as the probabilities are redistributed, and ultimately approaches the spin-temperature distribution $\rho= e^{\beta\hat{F}_z}/Z$, which is a NESS.
At that point, a balance is reached between the pumping ($R_{\text{op}}$) and relaxation ($\Gamma_{\text{SE}}$, $\Gamma_{\text{SD}}$) rates, such that the probabilities remain constant over time.
\begin{figure*}[t] 
    \centering 
    \begin{overpic}[width=0.98\textwidth]{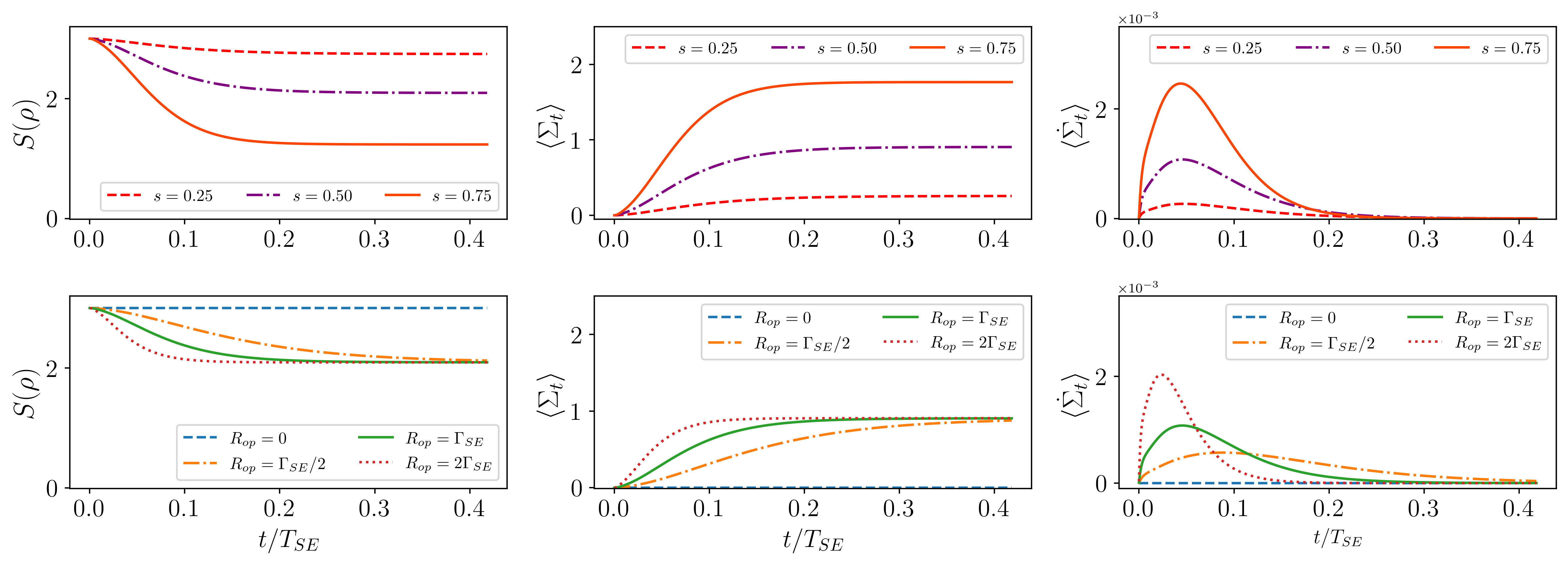} 
          \put(-0.5,34){(a)}
\put(33.5,34){(b)}
 \put(67.5,34){(c)}
 \put(-0.5,16.5){(d)}
 \put(33.5,16.5){(e)}
 \put(67.5,16.5){(f)}
    \end{overpic}
    \caption{Thermodynamic evolution of the atomic ensemble towards the steady state. The panels display the Von Neumann entropy $S(\rho)$, accumulated entropy production $\langle\Sigma_{t}\rangle$, and entropy production rate $\langle\dot{\Sigma}_{t}\rangle$ as a function of normalized time $t/T_{SE}$. The top row shows variations with light polarization $s$ setting $R_{op}=\Gamma_{SE}$, while the bottom row shows variations with pumping rates $R_{\text{op}}$ with $s=0.5$. The dynamics are solved for a vapor cell (radius $1.5\ \text{cm}$) at $120\ ^{\circ}\text{C}$ with buffer gas pressures of $200\ \text{Torr}$ He and $75\ \text{Torr}$ $\text{N}_2$. The relaxation rates used are $\Gamma_{\text{SD}} \approx 30\ \text{Hz}$ and $\Gamma_{\text{SE}} = 14\ \text{kHz}$. The monotonic decrease of $S(\rho)$ reflects the progressive spin ordering. The entropy production $\langle\Sigma_{t}\rangle$ saturates, quantifying the total irreversible cost of state preparation, while $\langle\dot{\Sigma}_{t}\rangle$ peaks during the transient regime and vanishes as the system reaches the NESS.}
\label{fig:Entropy}
\end{figure*}

In the context of magnetometry, researchers primarily focus on the dynamical aspects involved in magnetometer state preparation. Nevertheless, in the following section, we investigate the thermodynamic features of the evolution of the atomic-spin density matrix.

\section{Entropy and Irreversibility}~\label{secIII}
Among the processes involved in the magnetometer, optical pumping is inherently irreversible. This is evident as the pumping term is part of the incoherent dynamics (the dissipator) in Eq.~(\ref{eq:full_master_eq}). We can analyze the spin orientation process by examining the von Neumann entropy, $S(\rho)$. In turn, a reduction in this entropy indicates an increase in the system's information \cite{NielsenChuang}. A perfectly polarized state is a pure state characterized by $S(\rho)=0$ while a depolarized thermal state exhibits maximum entropy. Using $S(\rho_{t})$, it is possible to analyze the efficacy with which optical pumping orders the system against dissipative effects.

Another important process involves spin-exchange collisions. During the pumping process, the system is not isolated; it interacts with the laser and frequently collides with other atoms. These interactions cause relaxation and drive the system through irreversible processes. Therefore, entropy production is expected. We can quantify the total average entropy production as
\begin{equation}
\expval{\Sigma_{t}} =  D(\rho_{t} \| \rho_{\text{th}}),
\label{entropy_prod}
\end{equation}
where $D(\rho_{t}\|\rho_{\text{th}})=\Tr[\rho_{t}(\ln \rho_{t} - \ln \rho_{\text{th}})]$ corresponds to the relative entropy between the state $\rho_t$ and a reference thermal equilibrium state $\rho_{\text{th}}$ , which is the maximally mixed state\cite{ReviewLandi}.  Equation (\ref{entropy_prod}) characterizes the irreversibility of the dynamics~\cite{ReviewLandi,Deffner2011} and the rate of entropy production, $\langle\dot{\Sigma}_t\rangle=\frac{d}{dt} \langle\Sigma_t\rangle$, quantifies how quickly this irreversibility occurs. Therefore, a positive rate $\langle\dot{\Sigma_t}\rangle> 0$ corresponds to an ongoing irreversible process, while $\langle\dot{\Sigma}_{t}\rangle = 0$ indicates that the system has reached a steady state.

Figure~\ref{fig:Entropy} shows the results for the entropy, entropy production, and the rate of entropy production for different situations of the optical pumping process. First, we consider the system being pumped with polarized light along the $\hat{z}$-axis, as shown in the top row of Fig.~\ref{fig:Entropy}(a-c). For this scenario, we fix the pumping rate (e.g., ${R}_{\text{op}}=\Gamma_{SE}$) and vary the photon spin component $s$ for $\mathbf{s}=(0,0,s)$. This reflects realistic experimental conditions where light is not perfectly circularly polarized. As seen in Fig.~\ref{fig:Entropy}(a), a higher value of $s$ (more polarized light) drives the system to a lower steady-state von Neumann entropy $S(\rho)$, indicating a more ordered final state. This increased ordering comes at a thermodynamic cost, quantified by the entropy production $\langle\Sigma_{t}\rangle$ upon reaching the NESS. Figure~\ref{fig:Entropy}(b) shows that this cost increases with $s$, and the entropy production rate in Fig.~\ref{fig:Entropy}(c) reflects how irreversibility evolves during the dynamics. A larger value of $s$ leads to a more intense and rapid rate of entropy generation (a higher and earlier peak in $\langle\dot{\Sigma}_{t}\rangle$), which then correctly decays to zero as the system settles into its NESS. Nevertheless, the typical time at which the rate of entropy production reaches its maximum value, when the system becomes irreversible, is independent of the polarization. Therefore, the emergence of spin order renders the dynamics irreversible, transforming the system from a thermal equilibrium state to a polarized NESS.

 In the second situation, depicted in Figs.~\ref{fig:Entropy}(d-f), we maintain $\hat{z}$-polarized light with a fixed photon spin $s=0.5$ and vary the pump rate, $R_{\text{op}}$, which corresponds to varying the pump laser intensity. The plots show that a higher pump rate is more effective at ordering the system and drives it to the NESS faster than lower intensities. The results show that from $R_{\text{op}} = \Gamma_{\text{SE}}/2$ (orange dashed line) onward, the system reaches the same minimum asymptotic value for the von Neumann entropy $S(\rho)$. This behavior indicates a saturation of the atomic polarization: once the optical pumping is sufficient to counterbalance the relaxation processes, the NESS becomes independent of the laser intensity. Interestingly, Fig.~\ref{fig:Entropy}(e) reveals that this level of ordering is achieved at the same total cost of entropy production $\langle \Sigma_t \rangle$ with different features of irreversibility.  Figure ~\ref{fig:Entropy}(f) shows that a higher pump rate leads to a much larger and faster entropy production rate $\langle\dot{\Sigma_t}\rangle$, which reaches its peak earlier and drives the system more quickly toward its steady state. Therefore, the pumping rate determines both the timescale and the degree of irreversibility of atomic polarization.
 
It is worth noting that when atoms reach the NESS, they are prepared in a higher-energy state, i.e., most are in the $F=2$ state.  Thus, the pumping process polarizes the ensemble of atoms, raising their energy, which can then be used for spin precession in the magnetometer. The amount of work that can be extracted from this process is characterized by a thermodynamic variable, which can help define a figure of merit for its efficiency. 

\section{Ergotropy and polarization efficiency}~\label{secIV}
In conventional quantum thermodynamics, energy exchange is typically partitioned into work and heat, where work is associated with coherent changes in the Hamiltonian, while heat relates to changes in the population distribution of the energy eigenstates~\cite{Alicki1979,F.Nori_PRE06, dePaula2025}. However, for the driven-dissipative dynamics of our system, where the Hamiltonian remains time-independent, this standard distinction is not straightforward to apply.

Here, instead of focusing on the characterization of heat and work, we aim to quantify the amount of useful energy stored in the system during the preparation process. To this end, we employ the ergotropy measure, $\mathcal{E}(\rho)$ \cite{dePaula2025}, which is defined as the maximum amount of energy that can be converted into work via coherent, reversible dynamics~\cite{Allahverdyan_2004, Farina19, ReviewCampaioli, Carrasco22}, 
\begin{equation}
\mathcal{E}(\rho)= E(\rho) - \min_{\mathcal{U}}\text{Tr}\left[H_0\mathcal{U}\rho \mathcal{U}^{\dagger}\right],
\end{equation}
where the first term $E(\rho) = \Tr[H_0\rho]$ represents the internal energy, with Hamiltonian $H_{0} = \sum_{i} e_{i} |e_{i}\rangle \langle e_{i}|$ (with energy eigenvalues $\{ |e_{i}\rangle \}$ sorted in ascending order $e_{0} \leq e_{1} \leq ...$) and density operator $\rho = \sum_{i} r_{i} |r_{i}\rangle \langle r_{i}|$ (with populations $\{ |r_{i}\rangle \}$ sorted in descending order $r_{0} \geq r_{1} \geq ...$), respectively. The second term $E_{p} = \min_{\mathcal{U}}\text{Tr}\left[H_0\mathcal{U}\rho \mathcal{U}^{\dagger}\right]$ represents the energy of the passive state, $\rho_{p}=\sum_{i}r_{i}|e_{i}\rangle\langle e_{i}|$, unitarily connected to $\rho$ such that the highest probabilities are assigned to the lowest energy levels and thus no work can be extracted from this state. Therefore, we employ ergotropy to quantify the amount of useful energy available for spin precession, i.e., the maximum energy that atomic spin states can use to precess in a magnetic field.

In addition to ergotropy, we can define a figure of merit that measures the efficiency of that extractable work. Following the works of Refs.\cite{Farina19,Andolina19}, we quantify the polarization efficiency, $\mathcal{R}$, by means of the ratio between the extractable energy of the system (ergotropy) and its internal energy,
\begin{equation}
\mathcal{R} = \frac{\mathcal{E}(\rho)}{E},  
\end{equation}
where $0\leq\mathcal{R}\leq1$. Consequently, when $\mathcal{R}$ is close to 1, the polarized state has more energy available to do work, indicating a highly efficient state preparation process.

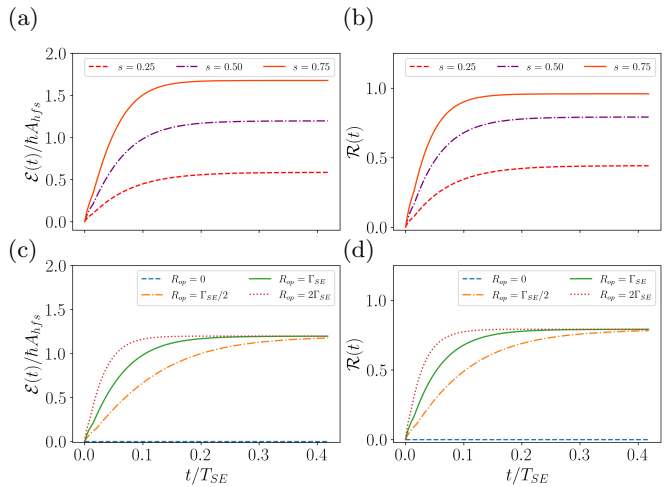
\begin{figure}[h]
    \centering
        \begin{overpic}[width=0.48\textwidth]{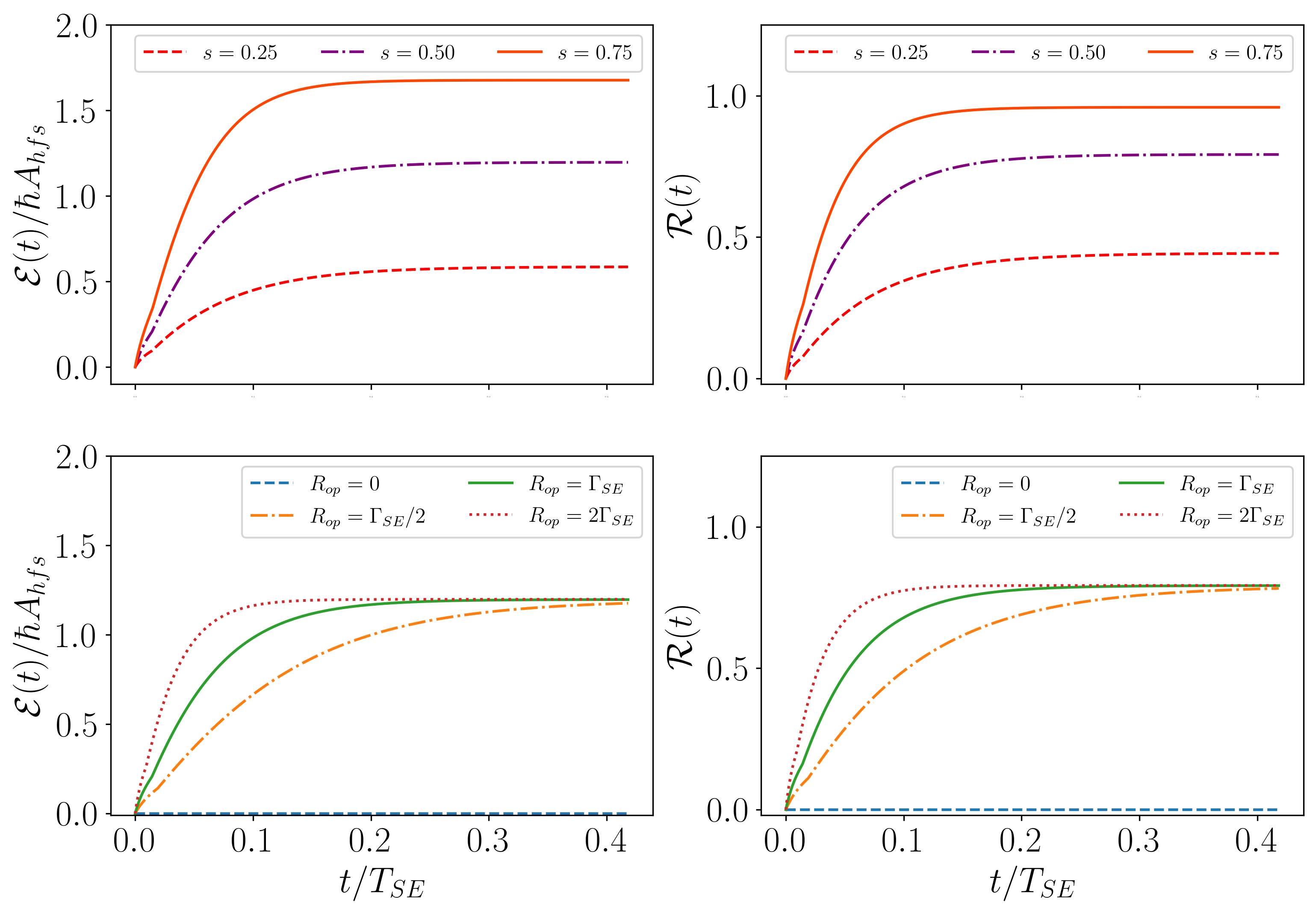} 
          \put(-1.7,72){(a)}
\put(50,72){(b)}
 \put(-1.7,36){(c)}
 \put(50,36){(d)}
\end{overpic}
    \caption{Ergotropy $\mathcal{E}(t)$ and the polarization efficiency $\mathcal{R}(t)$ as a function of time for different pumping rates $R_{\text{op}}$ and light polarizations $\vec{s}$. The top row shows variations with light polarization $s$ setting $R_{op}=\Gamma_{SE}$, while the bottom row shows variations with pumping rates $R_{\text{op}}$ with $s=0.5$. The dynamics are solved for a vapor cell (radius $1.5\ \text{cm}$) at $120\ ^{\circ}\text{C}$ with buffer gas pressures of $200\ \text{Torr}$ He and $75\ \text{Torr}$ $\text{N}_2$. The relaxation rates used are $\Gamma_{\text{SD}} \approx 30\ \text{Hz}$ and $\Gamma_{\text{SE}} = 14\ \text{kHz}$.}
    \label{fig:Ergotropy}
\end{figure}

In Fig.~(\ref{fig:Ergotropy}), we present the dynamic of both figures of merit $\mathcal{E}(\rho)$ and $\mathcal{R}$. We considered a situation of atomic density with $T=120^oC$ and the buffer pressure of $200\ \text{Torr}$ He and $75\ \text{Torr}$ $\text{N}_2$ Torr, with $\hat{z}$-polarized light, fixing the pumping rate at $R_{\text{op}}=\Gamma_{\text{SE}}$ for different values of the photon spin component $s$. As shown in Fig.~\ref{fig:Ergotropy}(a), the ergotropy always increases with time and converges to a value that depends on the amplitude of the photon spin vector $s$. As a result, Fig.~\ref{fig:Ergotropy}(b) demonstrates a higher polarization efficiency for more circularly polarized light wth photon spin of $s=0.75$, which limits the system to an efficiency of approximately $95\%$, while decreasing the polarization to $s=0.25$, the efficiency is reduced to $45\%$. Therefore, the light polarization controls the efficiency at which the atoms are optically pumped.

Now, for a particular case of light polarization, $s=0.5$, Fig.~\ref{fig:Ergotropy}(c) and (d)  show the ergotropy and polarization efficiency, respectively, for different values of optical pumping rate $R_{\text{op}}$. We observe that the system reaches the NESS faster for the highest rate ($R_{\text{op}}\geq\Gamma_{\text{SE}}$), achieving a polarization efficiency of approximately 80\%, whereas the efficiency drops to zero when the optical pumping rate is reduced up to $R_{\text{op}}=0$, corresponding to the pump being turned off.

\begin{figure}[t]
    \centering
    \includegraphics[width=0.9\linewidth]{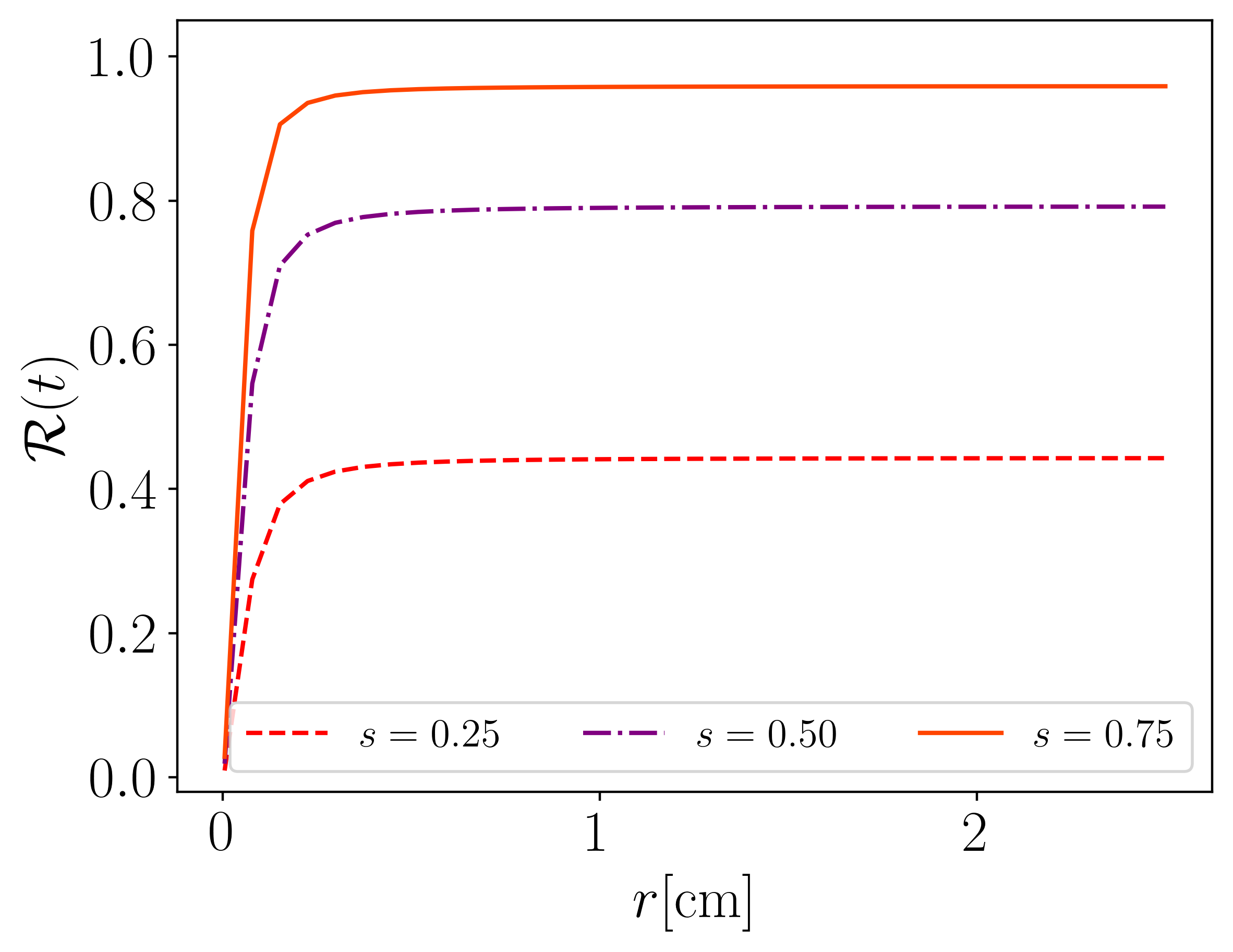}
    \caption{Polarization efficiency  $\mathcal{R}(t)$ as a function of the cell radius $r$. The system is evaluated at $T = 120$~$^{\circ}$C ($\Gamma_{SE} \approx 14$~kHz)  with buffer gas pressures of $200\ \text{Torr}$ He and $75\ \text{Torr}$ $\text{N}_2$, we fixed pumping rate $R_{op} = \Gamma_{SE}/2$. The plot shows the evolution for different $s$ values, where $\Gamma_{SD}$ varies from 21~Hz to 290~kHz as $r$ increases from 0.01~cm to 2.5~cm.}
    \label{efivsraio}
\end{figure}

Another aspect we analyze is the effect of spin destruction on polarization efficiency. To do so, we investigate the efficiency of the cell geometry, which can significantly change the spin description rate $\Gamma_{SD}$ as the cell size decreases. As the cell size decreases, wall collisions become more frequent. Figure~\ref{efivsraio} shows the steady-state polarization efficiency $\mathcal{R}(t)$ as a function of the cell radius $r$ for different photon spin values $s$. It is observed that for small radii (typically $r < 0.5$ cm), $\mathcal{R}(t)$ decreases abruptly, indicating that the depolarization rate induced by wall collisions dominates over the optical pumping rate. Conversely, for a larger radius, the efficiency saturates, suggesting that the system reaches its maximum theoretical polarization for a given $s$. This saturation shows that the gas buffer mitigates wall effects in very large cells, and the efficiency is primarily determined by the spin-exchange dynamics and the polarization of the light, making the steady-state performance independent of the cell geometry.

Therefore, as long as the spin exchange collision occurs at a higher rate than the spin destruction process ($\Gamma_{SE}> \Gamma_{SD}$), light can efficiently polarize dense atoms through an irreversible process,  as shown in Section~\ref{secIII}. The operational parameters that maximize polarization efficiency and ergotropy (specifically, high photon spin polarization $s$ and strong pumping rates $R_{\text{op}}$) correspond to the highest total entropy production $\langle\Sigma_t\rangle$ in Fig.~\ref{fig:Entropy}. This indicates that the storage of coherent energy in the atomic ensemble is associated with dissipative dynamics: attaining an ordered state requires driving the system far from equilibrium, thereby increasing entropy production through an irreversible process.

\section{Metrological Performance}\label{secV}

In addition to the thermodynamical properties of the preparation state through the pumping process described in the previous sections, the dynamics of eq.(\ref{eq:full_master_eq}) offer the possibility to analyze the metrological performance that can be exploited in the probing stage. Here, we study sensitivity limits through the QFI during the pumping process.

The connection between the QFI and the magnetic field arises from the parameter encoding driven by the Zeeman interaction~\cite{Budker2007, Degen2017,Giovannetti2006}. The Hamiltonian coupling the ensemble to the field $\mathbf{B}$ is $\hat{H} = -\gamma \mathbf{B} \cdot \mathbf{F}$. Thus, the atomic spin operator $\hat{F}_k$ acts as the generator of the unitary evolution that encodes the field $B_k$ into the quantum state.

The QFI defined as $\mathcal{F}_Q$ is a figure of merit for the metrological utility of the system, and it quantifies the intrinsic sensitivity of the state $\rho$ to changes in the parameter encoded by the generator $\hat{F}_k$\cite{Braunstein1994}. Considering the spectral decomposition of the density matrix, $\rho = \sum_i \lambda_i |\psi_i\rangle\langle \psi_i|$, where $\lambda_i$ and $|\psi_i\rangle$ are the eigenvalues and eigenvectors, respectively, the QFI becomes~\cite{Braunstein1994, Liu2014}:
\begin{equation}
\mathcal{F}_Q(\rho, \hat{F}_k) = 2 \sum_{i,j} \frac{(\lambda_i - \lambda_j)^2}{\lambda_i + \lambda_j} |\langle \psi_i | \hat{F}_k | \psi_j \rangle|^2,
\end{equation}
where the sum runs over indices for which $\lambda_i + \lambda_j > 0$. Physically, a larger $\mathcal{F}_Q$ indicates that the state reacts more strongly to the magnetic field interaction, carrying more information about the parameter $B_k$. According to the quantum Cramér-Rao Bound (QCRB)~\cite{Degen2017,Giovannetti2006}, the fundamental limit for the estimation variance of the magnetic field $B_k$ becomes $\Delta B_{k}\geq[\mathcal{F}_{Q}(\rho,\hat{F}_{k})]^{-1/2}$. In turn, a higher $\mathcal{F}_Q$ directly corresponds to a lower estimation error bound in the magnetic field.

\begin{figure}[t]
    \centering
    \begin{overpic}[width=1\linewidth]{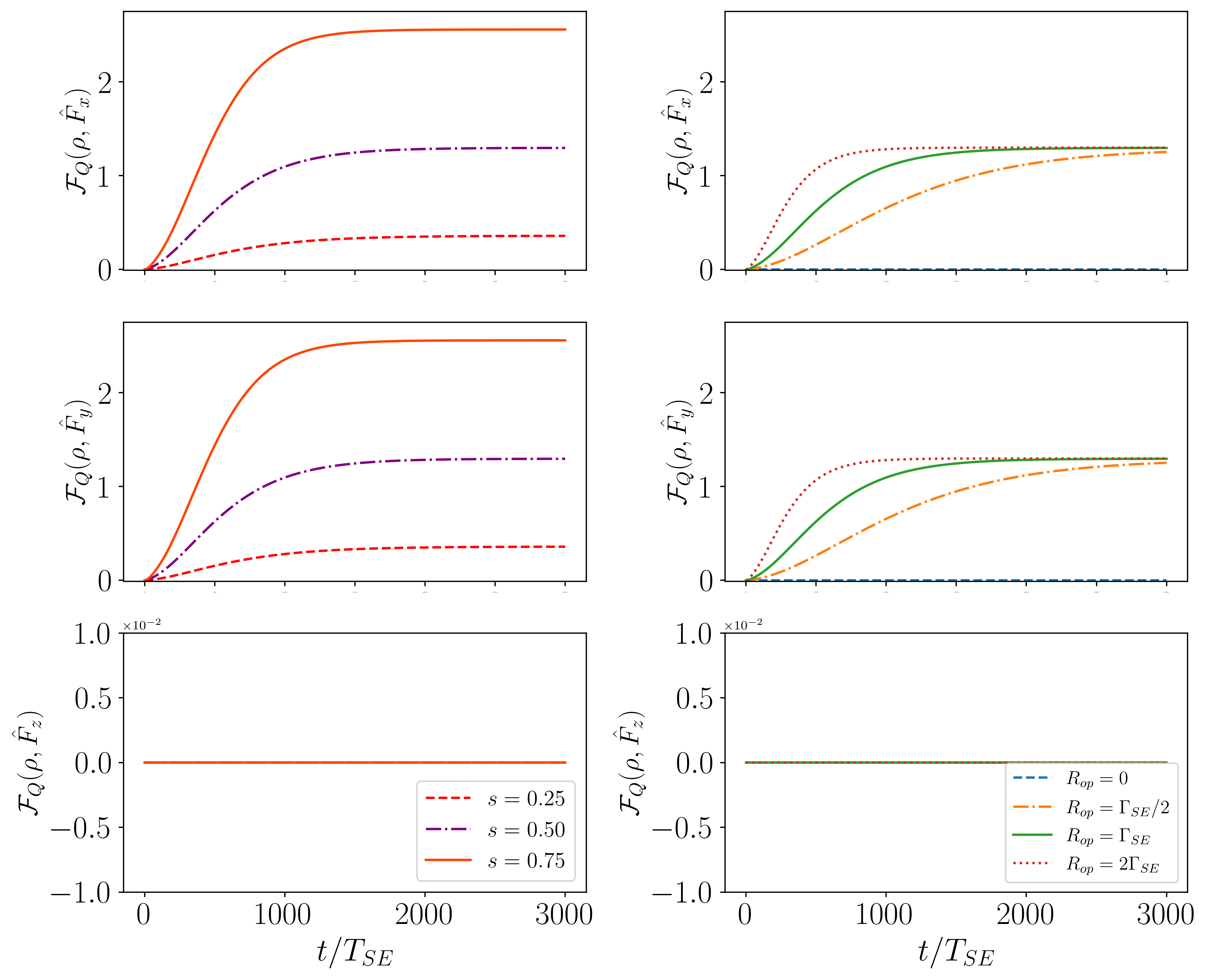}
        \put(0,80){(a)}
        \put(0,55){(b)}
        \put(0,28.5){(c)}
        \put(50,80){(d)}
        \put(50,55){(e)}
        \put(50,28.5){(f)}
    \end{overpic}
    \caption{Time evolution of the Quantum Fisher Information (QFI), $\mathcal{F}_Q(\rho,\hat{F}_k)$, calculated for the atomic spin operators $\hat{F}_x$, $\hat{F}_y$, and $\hat{F}_z$. The left column (a)-(c) shows the effect of varying the light polarization $s$ with a fixed pump rate ($R_{\text{op}}=\Gamma_{SE}$). The right column displays (d)-(f) the dynamics for fixed polarization ($s=0.5$) under different pump rates $R_{\text{op}}$. The dynamics are solved for a vapor cell (radius $1.5\ \text{cm}$) at $120\ ^{\circ}\text{C}$ with buffer gas pressures of $200\ \text{Torr}$ He and $75\ \text{Torr}$ $\text{N}_2$. The relaxation rates used are $\Gamma_{\text{SD}} \approx 30\ \text{Hz}$ and $\Gamma_{\text{SE}} = 14\ \text{kHz}$.}
    \label{fig:qfi}
\end{figure}

To analyze the metrological potential of the prepared states, we present the dynamics of the QFI in Fig.~\ref{fig:qfi} for the same parameters as in Figs.~\ref{fig:Ergotropy}(b) and (d). Figures~\ref{fig:qfi}(a-c) show the QFI for different values of the photon spin, whereas Fig.~\ref{fig:qfi}(d-f) presents the QFI for different pump rates. In the case of large values of $s$, which yields a higher polarisation efficiency in Figs.~\ref{fig:Ergotropy}(b) and (d) corresponds to the significantly larger values of   $\mathcal{F}_Q(\rho,\hat{F}_x)$ and $\mathcal{F}_Q(\rho,\hat{F}_y)$, as shown in Figs.~\ref{fig:qfi}(a) and (b). On the other hand, by increasing the pump rate up to $R_{op}\sim 2\Gamma_{SE}$, the QFI along the transverse directions reaches the same maximum limit, but at different time rates, as shown in Figs.~\ref{fig:qfi}(d) and (e). This indicates that reducing entropy and storing ergotropy effectively converts the ensemble into a resource for sensing transverse magnetic fields. Additionally, $\mathcal{F}_Q(\rho,\hat{F}_z)$ remains zero in all scenarios, as shown in Fig.~\ref{fig:qfi}(c) and (f). This occurs because the steady state is diagonal in the $\hat{z}$-basis and invariant under rotations about this axis. Consequently, the sensor is insensitive to field fluctuations parallel to the pump beam.

\begin{figure}[b!]
    \centering
    \begin{overpic}[width=1\linewidth]{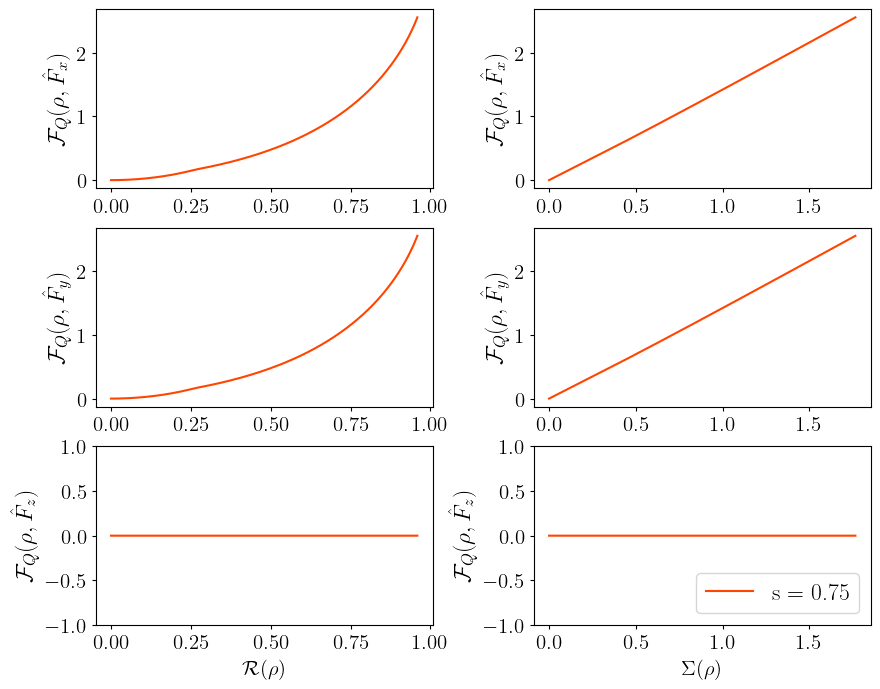}
        \put(0,80){(a)}
        \put(0,55){(b)}
        \put(0,28.5){(c)}
        \put(52,80){(d)}
        \put(52,55){(e)}
        \put(52,28.5){(f)}
    \end{overpic}
    \caption{Reparametrization of  $\mathcal{F}_Q(\rho,\hat{F}_k)$, as a function of the preparation efficiency $\mathcal{R}$ and the relative entropy $\Sigma(\rho)$. The left column (a)-(c) displays the nonlinear scaling of the QFI with respect to $\mathcal{R}(\rho)$, illustrating the accelerated gain in sensitivity as the state preparation approaches the maximum efficiency limit. The right column (d)-(f) reveals the strict linear proportionality between the QFI and $\Sigma(\rho)$, identifying the non-equilibrium information content as the primary driver for metrological precision. The curves are shown for a fixed light polarization $s=0.75$ and pump rate $R_{\text{op}}=\Gamma_{SE}$. The parameters correspond to a vapor cell at $120\ ^{\circ}\text{C}$ with $200\ \text{Torr}$ He and $75\ \text{Torr}$ $\text{N}_2$ ($\Gamma_{\text{SD}} \approx 30\ \text{Hz}$, $\Gamma_{\text{SE}} = 14\ \text{kHz}$).}
    \label{fig:qfi_ratio}
\end{figure}
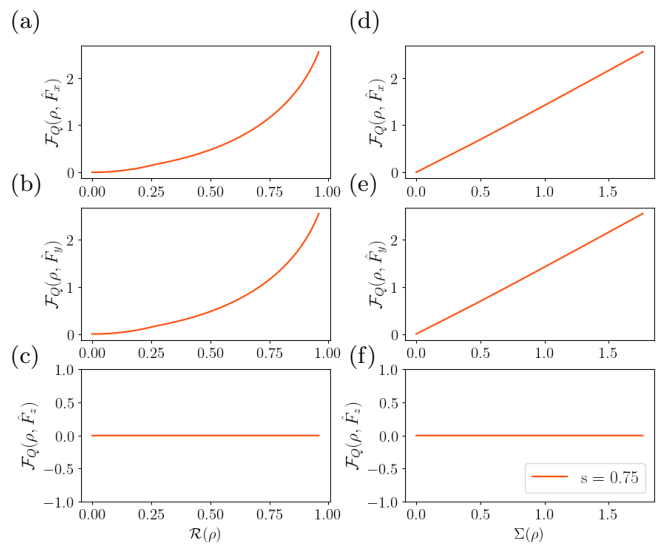

Complementing the temporal analysis, we investigate  $\mathcal{F}_Q(\rho,\hat{F}_k)$ as a function of the preparation efficiency $\mathcal{R}$ and the relative entropy $\Sigma(\rho)$. This reparametrization shifts the perspective from dynamical evolution to a resource-based representation, in which state trajectories $\rho(t)$ are evaluated in terms of their thermodynamic and informational content. 
As illustrated in Fig.\ref{fig:qfi_ratio} for $s=0.75$ and $R_{op}$ fixed (the regime of greatest polarization efficiency), the QFI exhibits distinct behaviors in each representation. We observe a non-linear scale with the preparation efficiency $\mathcal{R}$ in Figs.\ref{fig:qfi_ratio}(a) and (b), characterized by a superlinear growth of the QFI. This behavior indicates that the metrological response depends on the state preparation, with significant gains in QFI achieved only when the efficiency $\mathcal{R}$ approaches its maximum.
On the other hand, Figs.\ref{fig:qfi_ratio}(d) and (e) show the relationship between $\mathcal{F}_Q$ and the relative entropy $\Sigma$ is strictly linear. This proportionality reveals that the out-of-equilibrium information content directly affects performance. In this sense, each unit of informational deviation from the thermal state is converted at a constant rate in the metrological response. Figures~\ref {fig:qfi_ratio}(c) and (f)  $\mathcal{F}_Q(\rho,\hat{F}_z)$ remain zero throughout the evolution; thus, the sensor provides no metrological performance to field fluctuations in this direction.
\section{Conclusion}~\label{secVI}
In this work, we performed a detailed analysis of optical pumping from a thermodynamic perspective. This approach allows us to demonstrate how the pump rate and light polarization drive the system toward NESS. We analyzed these states by focusing on irreversibility and their potential useful energy. Our results highlight a trade-off: achieving higher useful energy incurs a higher thermodynamic cost.  Furthermore, we quantified the efficiency of the pumping process using ergotropy. In that context, we showed that higher polarization efficiency correlates with higher quantum Fisher information. This implies that a thermodynamically efficient preparation directly improves the metrological limits, lowering the bound on the magnetometer's sensitivity. The thermodynamic treatment presented in this work offers another approach to understanding optically pumped magnetometers. These results can also help identify which strategies to adopt to achieve a better-prepared state for magnetometer applications.

\section{Acknowledgments}
A.F.S. acknowledges the Coordenação de Aperfeiçoamento de Pessoal de Nível Superior – Brasil (CAPES) – Finance Code 001. C.H.S.V. acknowledges the São Paulo Research Foundation (FAPESP) Grant No. 2023/13362-0 and Grant No. 2025/14546-2 for financial support and the Southern University of Science and Technology (SUSTech) for providing the workspace during the internship. H.M.F. acknowledges the São Paulo Research Foundation (FAPESP) Grant No. 2024/18055-0, 2024/22385-6, and 2024/08522-0 for financial support. 

\appendix
\section{Simulation Parameters and Rate Estimation}
\label{apxA}

\begin{table}[h]
\centering
\caption{Spin-exchange and spin-destruction cross sections used in the calculations.}
\label{tab:cross_sections}
\begin{ruledtabular}
\begin{tabular}{c|c}
\textbf{Cross Section} & \textbf{Value (cm\textsuperscript{2})} \\
\hline
$\sigma_{\text{SE}}$               & $(1.9 \pm 0.1) \times 10^{-14}$ \cite{happer2010optically} \\
$\sigma_{\text{SD}}^{\text{Rb-Rb}}$ & $(9.0 \pm 0.5) \times 10^{-18}$ \cite{happer2010optically} \\
$\sigma_{\text{SD}}^{\text{Rb-He}}$ & $(8.7 \pm 0.5) \times 10^{-24}$ \cite{happer2010optically} \\
$\sigma_{\text{SD}}^{\text{Rb-N}_2}$ & $(1.0 \pm 0.1) \times 10^{-22}$ \cite{happer2010optically} \\
\end{tabular}
\end{ruledtabular}
\end{table}

For numerical solution, we consider a spherical hot vapor cell with 1.5cm of radius containing alkali metal (Rb$^{87}$), with He (200Torr) and N$_2$ (75Torr) buffer gas.

The Spin-Exchange Rate ($\Gamma_{\text{SE}}$), which describes the collisions between Rb-Rb atoms that conserve total spin, is given by: 
\begin{equation}
\Gamma_{\text{SE}} = n_{\text{Rb}} \cdot v_{\text{rel, RbRb}} \cdot \sigma_{\text{SE}},
\end{equation}
where $n_{\text{Rb}}$ is the atomic density of Rb, $v_{\text{rel, RbRb}}=\sqrt{16 k_B T / (\pi m_{\text{Rb}})}$  is the mean relative velocity for Rb-Rb collisions, and $\sigma_{\text{SE}}$ is the spin-exchange cross-section.

The total Spin-Destruction Rate ($\Gamma_{\text{SD}}$) accounts for all incoherent relaxation mechanisms that deplete the atomic spin polarization:
\begin{equation}
\Gamma_{\text{SD}} = \gamma_{\text{SD, RbRb}} + \gamma_{\text{SD, RbN}_2} + \gamma_{\text{SD, RbHe}} + \gamma_{\text{wall}},
\end{equation}
where $\gamma_{\text{SD, RbRb}} = n_{\text{Rb}} v_{\text{rel, RbRb}} \sigma_{\text{SD, RbRb}}$ represents the relaxation from spin-destroying Rb-Rb collisions. Similarly, $\gamma_{\text{SD, RbN}_2}$ and $\gamma_{\text{SD, RbHe}}$ describe the relaxation due to collisions with the buffer gases N$_2$ and He:
\begin{equation}
\gamma_{\text{SD, buffer}} = n_{\text{buffer}} \cdot v_{\text{rel}} \cdot \sigma_{\text{SD, buffer}}.
\end{equation}

The wall relaxation rate, $\gamma_{\text{wall}}$, arises from Rb atoms diffusing to and colliding with the cell boundaries. For cells with poor or no coatings, the accommodation coefficients $\alpha$ are not small enough to maintain unconfined polarization. Under these conditions of strongly depolarizing walls, the smallest spatial frequency of the diffusion mode is given by $k_1 \approx \pi/R$. Consequently, the relaxation rate due to the wall is expressed as:
\begin{equation}
\gamma_{\text{wall}} = k_D \cdot D(P),
\end{equation}
where $k_D = (\pi / R)^2$ is the diffusion mode constant for a radius sphere $R$. The term $D(P)$ is the effective diffusion coefficient of Rubidium in the buffer gas mixture, calculated as:
\begin{equation}
D(P) = \frac{D_0^{\text{He}}}{n_{\text{amg}}(P^{\text{He}})} + \frac{D_0^{\text{N}_2}}{n_{\text{amg}}(P^{\text{N}_2})},
\end{equation}
where $D_0^{\text{He}}$ and $D_0^{\text{N}_2}$ are the diffusion coefficients at reference conditions ($T_0 = 273.15$ K and $P_0 = 1$ atm), and $n_{\text{amg}}$ represents the gas density in Amagat units, referenced to 1 atm at 273.15 K.
The optical pumping rate ($R_{\text{op}}$) is not an intrinsic cell parameter but an external control parameter set by the laser to achieve a desired steady-state polarization. In the SERF regime, where $\Gamma_{\text{SE}} \gg \Gamma_{\text{SD}}$, we set $R_{\text{op}} \approx \Gamma_{\text{SE}}$.

\bibliography{ref}

\end{document}